\documentclass[aps,prl,reprint,superscriptaddress]{revtex4-1}

\usepackage{amsmath}
\usepackage{verbatim}
\usepackage{epsfig}

\newcommand{\om}{\omega}

\newcommand{\be}{\begin{equation}}
\newcommand{\ee}{\end{equation}}
\newcommand{\bea}{\begin{eqnarray}}
\newcommand{\eea}{\end{eqnarray}}

\begin{document}

\newcommand{\eg}{{\emph{e.g.} }}
\newcommand{\ie}{{\emph{i.e.} }}
\newcommand{\etal}{et al.}


\title{Force-Free Gravitational Redshift: Proposed Gravitational Aharonov-Bohm Experiment}


\author{Michael A. Hohensee}
\email{hohensee@berkeley.edu}
\author{Brian Estey}
\author{Paul Hamilton}
\affiliation{Department of Physics, University of California, Berkeley, CA 94720, USA}
\author{Anton Zeilinger}
\affiliation{University of Vienna and Institute of Quantum Optics and Quantum Information, Austrian Academy of Sciences, 1090 Wien, Austria}
\author{Holger M\"uller}
\affiliation{Department of Physics, University of California, Berkeley, CA 94720, USA}


\date{\today}

\begin{abstract}
We propose a feasible laboratory interferometry experiment with matter waves in  a gravitational potential caused by a pair of artificial field-generating masses. It will demonstrate that the presence of these masses (and, for moving atoms, time dilation) induces a phase shift, even if it does not cause any classical force. The phase shift is identical to that produced by the gravitational redshift (or time dilation) of clocks ticking at the atom's Compton frequency.  In analogy to the Aharonov-Bohm effect in electromagnetism, the quantum mechanical phase is a function of the gravitational potential and not the classical forces.
\end{abstract}


\maketitle

\begin{figure}
\centering
\epsfig{file=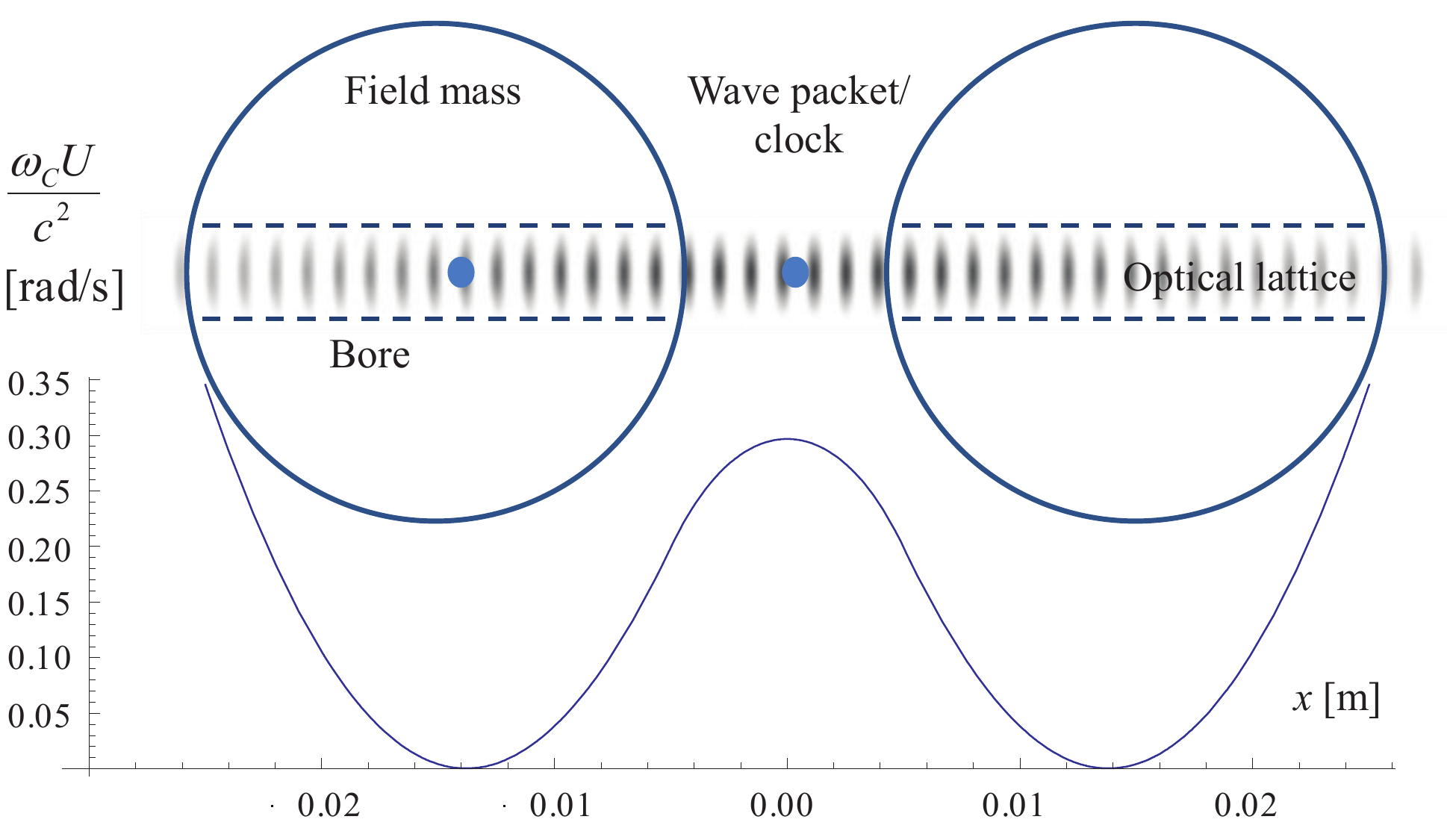,width=0.45\textwidth}
\caption{\label{setup} Setup. The source masses (radius $R=1$\,cm, density $\rho=10$\,g/cm$^3$) are separated by $L=3$\,cm. Wave packets are at saddle points of the potential $U(x)$, separated by $s=1.38$\,cm. The gravitational phase shift in rad/s is plotted for cesium atoms, for which $\om_C/(2\pi)=3\times 10^{25}$\,Hz. 
For $L=3R$, the gravitational potential difference is $\Delta U=1.11\rho G s^2$.
$L=2.61R, s=1.14R$ yields the largest $\Delta U$ for a given $s$, $\Delta U=1.17G\rho s^2$.}
\end{figure}

\begin{figure}[t]
\centering
\epsfig{file=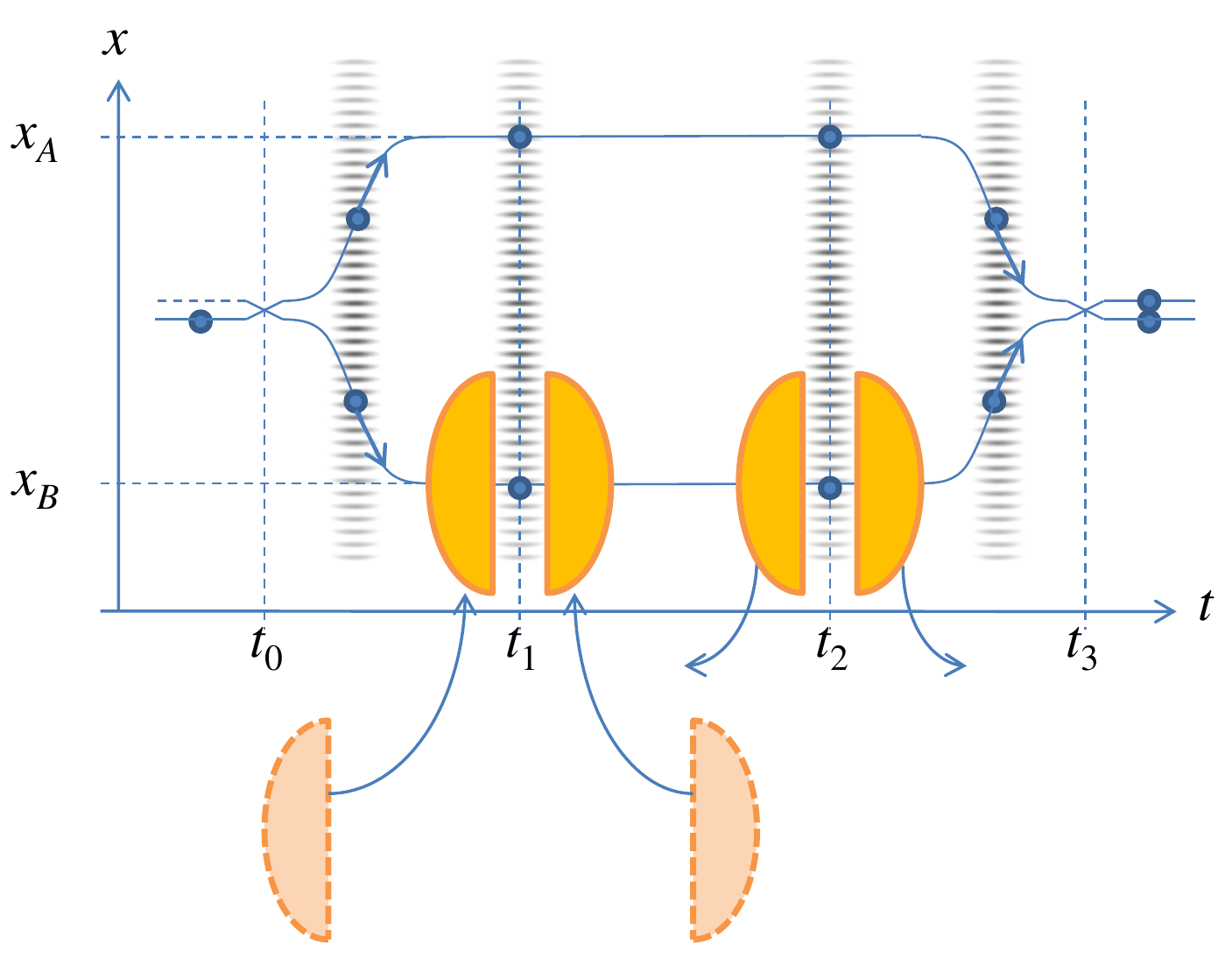,width=0.45\textwidth}
\caption{\label{schematic} Atom's trajectories versus time. The motion of one pair of test masses is sketched; the other pair (above $x_A$) is not shown.}
\end{figure}

The wave function of a particle in an interferometer
is measurably phase shifted by $\phi_A=-\frac e\hbar \int \vec A \cdot d\vec l$ or $\phi_V=\frac e\hbar \int V dt$ in the presence of a vector potential $\vec A$ or an electrostatic potential $V$, even in the absence of any classical force. This is the essence of the Aharonov-Bohm (AB) effect \cite{EhrenbergSiday1949,AharonovBohm1959,Batelaan}, which has been closely scrutinized and verified experimentally~\cite{ChambersMollenstedt,Caprez,Tonomura,Badurek}.  Gravitational analogs, broadly defined as phase shifts due to a gravitational potential $U$ in the absence of a gravitational force \cite{Sagnac,stringy}, have also been of great interest, but to date no experimental realization of a gravitational AB effect \cite{Ho1997,Zeilinger1983} has been suggested that would produce a signal of measurable size. Here, we suggest a feasible experiment (Figure \ref{setup}), using matter waves to probe the proper time in a multiply connected region of space-time comprised by two arms of an interferometer (Figure \ref{schematic}) in which the force caused by artificial gravitational field-generating masses vanishes. Using cold atoms, even the minuscule gravitational potential difference $\Delta U/c^2 \sim 1.6 \times 10^{-27}$ (Figure \ref{setup}) will produce a measurable phase difference \cite{redshift,redshiftPRL} $\phi_G=\om_C \int(\Delta U/c^2) dt$, owing to the long ($\sim 1\,$s) coherence times possible in such a system, and the large value of the atom's Compton frequency, $\om_C=mc^2/\hbar$. 
This phase shift is identical to that which accumulates between two clocks oscillating at $\om_{C}$ that record the proper time along each arm of the interferometer, and is naturally described as such in the context of general relativity. The experiment can also measure the time-dilation phase $\phi_T=-\frac12 \om_C\int \Delta(v^2)/c^2\,dt$ of moving wave packets, similar to what has been demonstrated for trapped ion clocks~\cite{Chou}. It will thus show that matter-waves indeed accumulate phase at the Compton frequency, modified by the local gravitational potential and time dilation, rather than simply moving in response to the local gravitational acceleration.

The phase measured by Mach-Zehnder matter-wave interferometers built so far ({\em e.g.,} \cite{Colella,KasevichChu,PritchardReview}) can be described by three effects - the wave packet's integrated gravitational redshift $\phi_G$, time dilation $\phi_T$, and interactions with the diffraction gratings ({\em e.g.}, standing light waves or a crystal). These effects are related to one another as $1:-1:1$, and only their sum is observed \cite{redshift,redshiftPRL,Wolf,Samuel,Giulini}. Thus, it is possible to view the interferometer as a measurement of the gravitational redshift caused by the gravitational potential between two Compton frequency oscillators \cite{redshift,redshiftPRL}, or to ignore the Compton frequency dynamics of the wavepacket, and even deny its physical relevance \cite{Wolf,Samuel,Giulini}. In the latter picture, the measured phase is ascribed to the phase of the gratings at the positions of the wave packets when interacting; the interferometer is thus argued to be a measurement exclusively of the wave packet's acceleration of free fall $\vec g$ in response to the gravitational force. These two views mirror the discussion on the influence of forces and potentials in quantum mechanics which surrounded the original Aharonov-Bohm paper \cite{AharonovBohm1959}. 
The gravitational AB effect arises from $\phi_G$ in the absence of a gravitational acceleration $\vec g$ and will conclusively show that the influence of the gravitational potential in quantum mechanics is physically relevant, measurable, and needed to describe observations. It is topological~\cite{Peshkin}, {\em i.e.} impossible to ascribe to any local effect such as a grating's phase at certain locations, and nondispersive~\cite{Badurek}, {\em i.e.,} not arising from motion or distortion of the wave packets.  The influence of the gravitational potential is identical to the gravitational redshift \cite{redshift,redshiftPRL}; interpretations that discount this are incomplete.

Figure \ref{setup} shows two identical spheres whose combined gravitational potential has a saddle point 
between the spheres ($x_A=0$) and another, lower potential saddle point at $\pm x_B$ close to the individual spheres' centers. An interferometer is formed by placing an atom of mass $m$ into a superposition of two quantum states at a time $t_0$ (Figure \ref{schematic}). These states are then conveyed to the two saddle points by moving optical lattices and held there for a time $T=t_2-t_1,$ during which they accumulate phase shifts $\phi_A, \phi_B$. Interferometers making similar use of optical lattices have already been demonstrated experimentally \cite{BBB,Charriere}. When the states are interfered at $t_3$, the phase difference $\Delta \phi=\phi_A-\phi_B$ can be measured by detecting the population in the outputs of the interferometer, which is given by $\cos^2 \Delta \phi/2$. The gravitationally induced phase difference produced by the source masses is $\delta \phi_G=m \Delta U T/\hbar$, and is analogous to the electrostatic AB effect \cite{Batelaan}. For a given separation $s$ between the potential maximum and minimum along the axis separating the masses, with $L=3R$ and $R=0.72 s$, the gravitational phase shift is
\be\label{deltaphiG}
\delta \phi_G=0.16\left(\frac{s}{\rm cm}\right)^2\left(\frac{\rho}{10\,  {\rm g/cm}^3}\right)\left(\frac{m}{m_{\rm Cs}}\right)\left(\frac{T}{{\rm s}}\right),
\ee
where $m_{\rm Cs}$ the mass of Cs atoms.

To derive this phase shift, consider an experiment without the source masses. We assume that the wave packets $|\psi_A(t_1)\rangle$ and $|\psi_B(t_1)\rangle$ at $t_1$  are concentrated near $x_A$ and $x_B$, respectively (Figure \ref{schematic}) and are eigenstates of a Hamiltonian $H_{A,B}$ describing all relevant potentials, in particular the optical lattice and Earth's gravity. Since the Hamiltonian is time-independent between $t_1$ and $t_2$, the time evolution is simply
\be
|\psi_{A,B}(t_2)\rangle=e^{\frac i\hbar H_{A,B} T}|\psi_{A,B}(t_1)\rangle .\label{eqtwo}
\ee
When the states interfere with each other at $t_3$, a phase difference $\phi_0$ is measured that results from the combination of all influences on the atom between $t_0$ and $t_3$.

Now consider the experiment with the source masses being brought in 
at $t_1$ and removed at $t_2$ (for simplicity, we neglect the time these processes require, although as we shall see, this need not be assumed to demonstrate the gravitostatic AB effect). Once in place, the masses apply no potential gradients or forces to the wave packets, provided that the wave packet is much smaller than the masses. They will thus not change the shape or location of the wave packets. The time evolution is now
\be
|\psi_{A,B}(t_2)\rangle= e^{\frac i\hbar [H_{A,B} T+mU(x_{A,B})T]}|\psi_{A,B}(t_1)\rangle.
\ee
If these states are interfered at $t_3$, they will have picked up an additional relative phase
\be\label{phiG}
\phi_G=m\Delta U T/\hbar.
\ee
To illustrate the utility of a matter wave packet as a clock in general relativity, we recall that the proper time experienced by a clock moving with velocity $v$ at a location $x(t)$ relative to a resting observer at $x=0$ is given by (see, {\em e.g.}, \cite{redshiftPRL})
\be\label{proptime}
\tau=\int\left(\frac{U(x)-U(0)}{c^2}-\frac 12 \frac{v^2}{c^2}\right)dt,
\ee
to leading order in the gravitational potential $U$ and the velocity $v$. We now consider a pair of clocks ticking at a proper frequency of $\om$ that are moved along the atom's paths. They are initially synchronized at $t_0$ and are compared at $t_3$. The clocks register a proper time difference
\be
\Delta \tau_0=\frac{1}{c^2}\int [U_0(x_A)-U_0(x_B)] dt,
\ee
where the kinetic term vanishes because of the symmetry of the trajectories. The notation $\tau_0, U_0$ indicates that these quantities are measured in the absence of the source masses. Adding the source masses changes the proper time difference by $\Delta \tau_G=\Delta U T/c^2$. (As above, $\Delta U$ refers to the potential difference caused by the source masses between the clock's locations.) For a clock frequency $\om$, the source masses give rise to an additional phase shift
\be
\Delta \phi_G=\om\Delta \tau_G= \om \Delta U T/c^2.
\ee
This phase shift is identical to the phase shift (given in Eq.~(\ref{phiG})) that the masses induce on matter waves, provided that we substitute $\omega=\omega_{C}$.  
At this, and only this frequency, a clock will acquire the same phase as a matter wave, and will do so in any gravitational potential, moving along any trajectory.  This is no coincidence: the expressions (2-4) follow from a path integral $\int\mathcal Dq\, e^{iS(q,\dot{q})/\hbar}$, where 
\begin{equation}\label{eq:rel}
S/\hbar=\frac{mc^{2}}{\hbar}\int d\tau=\omega_{C}\int \frac{1}{c}\sqrt{-g_{\mu\nu}dx^{\mu}dx^{\nu}}.
\end{equation}
All gravitational effects are described by the dimensionless metric tensor $g_{\mu\nu}=\eta_{\mu\nu}+h_{\mu\nu}$, where $\eta_{\mu\nu}$ is the Minkowski metric, and $h_{\mu\nu}$ is the dimensionless gravitational strain tensor.  
The weak gravitational field (with Newtonian potential $U$) enters the above expression for $d\tau$ via  $h_{00}=2U/c^{2}$, and so any local description of the gravitationally induced phase shift of a massive particle must be proportional to $\om_{C}h_{00}dt$. 
The Hamiltonian formulation leading to Eq. (\ref{phiG}) has been shown to be equivalent to the relativistic dynamics of matter waves, oscillating at $\om_C$~\cite{deBroglie}, propagating in curved space-time in the weak field limit (see Appendix A in ref. \cite{PQE}).  The non-relativistic treatment can also be used to derive the same interferometer phase difference, and would attribute it to the product of the atoms' mass with a dimensionful Newtonian potential.  The relativistic theory (Eq.~\ref{eq:rel}), however, describes matter waves as clocks that tick at the atoms' Compton frequency \cite{redshift,redshiftPRL}.

Were this force-free gravitational redshift, or gravitostatic AB effect, to be measured by atomic clocks, it would require km-sized source masses. Alternatively, clocks could be located at different Lagrange points of the Earth-Moon system.  Laboratory-scale tests, however, can make use of matter-wave clocks.  

Since we cannot turn off 
Earth's gravity, a true type I AB test (characterized by the complete elimination of {\em any} force acting on the wave packet~\cite{Batelaan}) would only be realizable in microgravity.  Nevertheless, as derived above in Eqs. (\ref{eqtwo}-\ref{phiG}), such an experiment can be approximated in the laboratory using an apparatus to move the source masses into place after the wavepackets have reached their respective holding positions $x_{A}$ and $x_{B}$, with the masses' trajectories selected such that they produce no significant forces at any time.  The effect of Earth's gravity can then be suppressed by comparing measurements made with and without the source masses.

This gravitostatic AB effect 
is both non-dispersive \cite{Badurek} and topological \cite{Peshkin}. 
The latter follows immediately from the fact that the interferometer phase 
is proportional to the line integral of a gauge-dependent integrand~\cite{Peshkin} (here, the local gravitational potential).  
The sources' force-free configuration makes this even more obvious: no gravimeters confined to the neighborhood of $x_{A}$ and $x_{B}$ could register the masses' presence.

Although the linear gradient of the field masses' potential vanishes at the saddle points, the curvature of their potential can modify the quantum states of the atoms to produce a dispersive phase shift.  This, however, is negligible. The dynamics of the atoms in the optical lattice potential $V$ can be approximated by a three-dimensional harmonic oscillator with eigenfrequencies $\om_i^2=(1/m) \partial^2 V/(\partial x_i^2)$, where $x_i=x,y,$ or $z$. The time evolution of the eigenstates is given by $e^{iE_h t/\hbar}$, where $E_h=\sum_i \hbar \om_i/2$. Adding the field masses modifies the potential $V\rightarrow mU+V$, and thus shifts the $\om_i$ by $\Delta \om_i =1/(2\om_i)\partial^2 U/(\partial x_i^2)$
to first order. This causes a corresponding modification of the time evolution. We note that this may vanish identically, if all $\om_i$ equal one another, by virtue of the Laplace equation $\nabla^2 U=0$. Otherwise, the change in the time evolution phase is on the order of
\be
\phi_{\rm curv} =\frac 23 \frac{\pi G \rho}{\omega_i}\sim 2\times 10^{-6}\,{\rm rad/s},\label{eq:quad}
\ee
where the lowest (radial) $\omega_i/2 \pi\sim 0.1$\,Hz was inserted for a conservative estimate. This is negligible; it can also be quantified and removed by varying the lattice depth, and thus the $\om_i$.

Additional dispersive phase shifts could arise from forces acting on the atoms. Due to the optical lattice, a small residual force $F$ will not cause a permanent velocity change, but only a shift $\delta x$ in the expectation value of the atoms' position. 
If the lattice potential is $-V_0\cos^2 k x$, the shift amounts to $\delta x=F/(2 k^2 V_0)$, and the resulting potential change of $F^2/(4 k^2 V_0)$ causes a negligible phase shift of
\be\label{phiF}
\phi_F=\frac{F^2 T}{4 k^2 V_0\hbar}.
\ee
The dispersive phase due to earth's gravitational force (Table \ref{tab:systematics}, line 6) can be suppressed by comparing the experiment with and without field-generating masses. The phase shift induced by the uncompensated force resulting from just one of the source masses (line 8) is negligible.  Dispersive and force-related phase shifts produced while the source masses and/or the wave packets are in motion are negligible for similar reasons. Experimentally, this can be verified by varying the time $T$ (Figure \ref{schematic}) 
while all other experimental parameters are kept constant.  Provided that any systematics produced by the source masses themselves are sufficiently small, this also means we can use fixed source masses (\ie a type II test), simplifying the experiment.

In type II electrostatic AB experiments \cite{MatteucciPozzi}, the electrons encounter a nonvanishing force on part of their way. This  modifies the time spent on the trajectories and could mimic the AB effect \cite{Boyer}. Our gravitostatic tests (type I and II) are free of this loophole due to the trapping action of the optical lattices. Gravitomagnetic forces caused by to the motion of the source masses, too, can be neglected, as they are suppressed by at least one power of their velocity over $c$. Finally, the source masses' gravitational potential will phase-shift the laser beams forming the optical lattice. The light cone $g_{\mu\nu}x^\mu x^\nu=0$ is given by $g_{tt}\simeq 1-2 U/c^2$ and $g_{xx}\simeq -(1+2 U/c^2)$, where $U$ is given both by Earth's field and the source masses. Thus, the source masses will fractionally shift the lattice at the order of $\Delta U/c^2$, which is smaller than $10^{-25}$, producing negligible phase shifts. 

\begin{table}
\caption{\label{tab:systematics} Contributions to the signal. Quantities marked (*) are common to each arm of the interferometer and cancel out; those marked (**) are independent of the source masses and can be removed by running with and without the source masses. We assume the dimensions stated in the caption of Figure \ref{setup} and in the text, and $T=1\,s$.}
\begin{tabular}{lllr}\hline\hline
& Source & & phase (rad) \\ \hline
1 & {Gravitostatic AB} & {Eq. (\ref{deltaphiG})} & ${0.3}$\\
2 & Earth's gravity** & $gs \omega_C T/c^2$ &  $2.8\times 10^{8}$ \\
3 & Lattice Shift* & $V_{0} T/\hbar$ & $ 6\times 10^{5}$\\
4 & Differential Lattice Shift** & See text & $0\pm 0.02$ \\
5 & Mean Field** & $4\pi \hbar a (\Delta n)T/m$ & 0.03 \\
6 & Dispersive (Earth's gravity)* & Eq. (\ref{phiF}) & $0.26$\\
7 & Quadratic Potential Shift & Eq.~(\ref{eq:quad}) & $ 2\times 10^{-6}$\\
8 & Dispersive (field mass) & Eq. (\ref{phiF}) & $2\times 10^{-8}$\\
9 & Magnetic Fields ($1$ mG) & $\frac{430 \text{Hz}}{G^{2}}(\Delta B)^{2}T$ &  $2\times 10^{-5}$ \\
\hline\hline
 \end{tabular}
 \end{table}

Verification of Eq. (\ref{deltaphiG}) at the level of 10 standard deviations will require systematic and technical errors to be below about 30\,mrad. Table \ref{tab:systematics} shows a number of potential systematic effects. We assume that the  $x$-axis is vertical. The leading two contributions to the background phase $\phi_0=gs\om_C T/c^2$ are Earth's gravity (line 2), and the optical lattice potential. The latter is mostly common to both arms of the interferometer (line 3), but a differential shift of $-2V_{0} T x_w s/(z_R^2 \hbar)$, where $z_R=\pi w_0^2/\lambda$ is the beam's Rayleigh range, remains due to the diffraction of the Gaussian beam, if the lattice beam waist is located at $x_w\neq 0$. We assume $V_0/h=100\,$kHz, $x_w=0\pm 1$\,mm, an $1/e^2$ intensity radius of $w_0=0.5\,$mm, and $\lambda=852\,$nm (line 4). The mean field shift due to atom-atom interactions produces a measurable phase shift proportional to the difference $\Delta n$ between the atomic densities in the interferometer arms (line 5). We assume $n=2\times10^{9}\,$cm$^{-3}$, $\Delta n/n\leq 0.016$, and a scattering length $a=3000 a_B$, where $a_{B}\equiv 5.3\times 10^{-11}$ is the Bohr radius. Effects that are independent of the source masses can be suppressed by comparing the phase with and without source masses. The required stability and resolution of parts in $10^9$ has already been demonstrated in atom interferometers \cite{Petersmetrologia,LVGrav}. If needed, further suppression can be achieved with paired co-interferometers \cite{McGuirk}: one interferometer measures the gravitational phase due to the source masses while a second, without source masses, simultaneously measures the background phase for subtraction.  If controlled by the same laser beams, the two interferometers can be identical to high precision. An effect correlated to the source masses arises from Zeeman shifts due to residual magnetism of the masses (Table \ref{tab:systematics}, line 9). To minimize it, we choose $m_F=0$ quantum states. 
Since residual iron content of the source masses may be suppressed to the level of parts per million, they will not be ferromagnetic. Magnetic shielding can be enhanced by a thin tube of mu metal (also serving as an electrostatic shield), and can be used to shield the entire apparatus.

The experiment can also measure the time dilating effects due to relative motion (the second term in Eq. (\ref{proptime})) on the phase of a matter wave clock~\cite{deBroglie}. This is accomplished by moving one of the wave packets periodically, so that $x(t)\rightarrow x_{0}(t)+A\sin \om t$, where $x_{0}(t)$ is the atoms' equilibrium position, $A$ an amplitude, and $\om$ a frequency.  This will produce a total phase shift of $\om_C \bar{v^2}T''/(2c^2)=\om_C A^2\om^2T''/(4c^2)$, where $T''$ is the total  duration of the oscillation. For $A= 0.1\,\mu$m, $\om/(2\pi)=1$\,kHz, and Cs atoms, this phase is 207\,rad per second. This oscillation can be induced by adiabatically shaking one of the optical lattices while the atoms are in motion (Figure \ref{schematic}); between $t_2$ and $t_3$, the two optical lattices become degenerate and it will not be possible to shake one wave packet without shaking the other. Alternatively, causing the two wave packets to have unequal laboratory-frame velocities while in motion will also produce such a time dilation phase. (As before, the fact that this phase can be derived from the Schr\"odinger equation is compatible with the fully relativistic picture \cite{deBroglie,PQE}.) By separately confirming the time dilation and gravitational redshift effects, this test will firmly establish the equivalence between matter waves and clocks.

The gravitostatic AB effect considered here requires that there be no classical forces acting on the atoms, which is equivalent to vanishing Christoffel symbols in the atom's rest frame. Other definitions~\cite{Dowker1967,strings,combinations} go further and require a vanishing Riemann tensor.  
Since the Riemann tensor does not vanish in our experiment, rapidly moving particles may still feel a force, though the force acting on the atoms at rest is zero. 

We have described an experiment that can unambiguously demonstrate that matter-waves are clocks that can be used to measure the gravitational redshift caused by the gravitational potential, and not merely rocks that provide quantum measurements of the gravitational acceleration: a matter-wave is subject to the same gravitational redshift and time-dilation effects that apply to a conventional clock, even if it is constrained to a space time region of vanishing gravitational force. This will be the first demonstration of a force-free gravitational redshift, and the first experimental demonstration of a gravitostatic Aharonov-Bohm effect.  The effect is non-dispersive and topological, and thus impossible to ascribe to any local influences on the wave packet.  While the proposed experiment would benefit greatly from being performed in microgravity, as this would strongly suppress the background phase due to
the Earth's field and permit increased interaction times,
it can nevertheless be realized in a terrestrial laboratory. 
The experiment will separately demonstrate the effect of time-dilation on matter-wave clocks.

We are indebted to S. Chu and M. Peshkin for important discussions, to J. M. Brown for careful reading of the manuscript, and the David and Lucile Packard Foundation, the National Aeronautics and Space Administration, the National Institute of Standards and Technology, the National Science Foundation, and the Alfred P. Sloan Foundation for support.



\end{document}